\documentstyle[12pt]{article}    
\begin{document}           
\rm
\baselineskip=0.33333in
\begin{quote} \raggedleft TAUP 2543-98
\end{quote}
\vglue 0.4in
\begin{center}{\bf Interrelations Between the Neutron's Magnetic \\
Interactions and the Magnetic Aharonov-Bohm Effect}
\end{center}
\begin{center}E. Comay 
\end{center}
 
\begin{center}
School of Physics and Astronomy \\
Raymond and Beverly Sackler Faculty \\
of Exact Sciences \\
Tel Aviv University \\
Tel Aviv 69978 \\
Israel
\end{center}
\vglue 0.3in
\noindent
Email: eli@tauphy.tau.ac.il
\vglue 0.3in
\noindent

\vglue 0.3in
\noindent
Abstract:

It is proved that the phase shift of a polarized neutron interacting
with a spatially uniform time-dependent magnetic field, demonstrates
the same physical principles as the magnetic Aharonov-Bohm effect.
The crucial role of inert objects is explained,
thereby proving the quantum mechanical nature of the effect. 
It is also proved that the nonsimply connectedness of the field-free 
region is not a profound property of the system and that it
cannot be regarded as a sufficient condition for a
nonzero phase shift.

\newpage
{\bf 1. Introduction}
\vglue 0.33333in

The interaction of the neutron's magnetic moment with external 
magnetic field has been used for studying properties of magnetic
materials, for understanding
the nature of the neutron's magnetic moment[1-3]
and for an analysis of general physical principles[4,5].
In [5], an examination is carried out for
a polarized neutron travelling through
a time dependent, spatially uniform magnetic field 
of a solenoid. The field is
parallel to the neutron's magnetic dipole
and to its velocity. These properties
provide an environment where the
neutron travels through a force-free and torque-free
region. Thus, taking a classical point of view, one may conclude
that the neutron can be regarded as a free particle.
Nevertheless, the experiment shows that the neutron
acquires a phase shift which
affects its interference pattern. 

   The authors of [5] relate their experiment to a kind of the
Aharonov-Bohm (AB) effect[6,7]. 
The fact that the neutron behaves as an inert object
and remains in its quantum
mechanical ground state has been 
analyzed recently[8]. It is proved there that this
property is essential for the phase shift obtained.
Indeed, if the neutron is replaced by a ``classical neutron", then the
phase shift disappears.

   The present work performs a further analysis of the neutron
experiment [5]. 
It is proved that this 
experiment demonstrates the same physical principles as
the electron interference experiment[9] showing the existence of
the {\em magnetic} AB effect.

   The discussion carried out below assumes that
the nonrelativistic limit holds. 
Units where $\hbar=c=1$ are used. 

   The physical elements of the polarized neutron experiment
are summarized in section 2. A description of the magnetic
AB effect is presented in section 3. In section 4 it is proved
that the two experiments demonstrate the same physical principles.
In section 5 it is proved that topological features of
the field-free region cannot be regarded as a necessary and
sufficient condition for a nonzero phase shift.
Concluding remarks are the contents of the last section.

\vglue 0.666666in
\noindent
{\bf 2. The Polarized Neutron Experiment}
\vglue 0.33333in

   Let us examine the physical properties of the neutron experiment[5].
The neutron's nonrelativistic Lagrangian boils down to
the following expression[8]
\begin{equation}
L = \frac{1}{2}mv^2 + \mbox{\boldmath $m \cdot B$}
\label{eq:LAGN}
\end{equation}
where {\boldmath $m$} denotes the neutron's magnetic moment and
{\boldmath $B$} is the magnetic field of the solenoid.
The first term of this Lagrangian is independent of the external
magnetic field {\boldmath $B$} (because the neutron 
travels in a force-free and torque-free region) whereas the
second term is proportional to this field. Therefore,
the action and its associated phase vary due to the interaction
of the external magnetic field {\boldmath $B$}
with the neutron's magnetic moment {\boldmath $m$}.

   Evidently, the moving neutron is a part of the entire system.
The full picture is obtained after including the solenoid's
interaction with the moving neutron. The magnetic field
{\boldmath $B$} of $(\!\!~\ref{eq:LAGN})$ is associated
with the motion of charges along the solenoid's wires.
Thus, $(\!\!~\ref{eq:LAGN})$ is a part of a system which is
a sum of 2-body interactions of the following kind
\begin{equation}
L = L_n + L_e + L_{ne},
\label{eq:LLL}
\end{equation}
where $L_n$ and $L_e$ denote the single particle interactions of the
neutron and the electron, respectively and $L_{ne}$ denotes the
neutron-electron interaction. The summation of $L_{ne}$
of $(\!\!~\ref{eq:LLL})$ on all charges boils down to the
last term of $(\!\!~\ref{eq:LAGN})$.

   Obviously, the 2-body Lagrangian $(\!\!~\ref{eq:LAGN})$ yields
a 2-body Hamiltonian whose Schrodinger equation that takes the form
\begin{equation}
(H_n + H_e + H_{ne})\psi = i\hbar \frac{\partial \psi }{\partial t}.
\label{eq:SCH}
\end{equation}

   In the neutron experiment [5], it is assumed that the neutron
does not affect the solenoid's state. The correctness of this
assumption is proved here. Performing a Lorentz
transformation on the magnetic field of a motionless neutron,
one realizes that in the laboratory frame, the moving neutron has
an electric field that acts on
the solenoid's current. However, it is easy to see
that the solenoid's current is not affected by this field.
This point, which is generally taken for granted,
is proved here because of its
importance for the following discussion.

   The overall EMF force of the electric field of the moving
neutron is obtained by integrating this field along the
solenoid's wires. The magnetic field of a magnetic dipole
is[10]
\begin{equation}
\mbox{\boldmath $B$} = [3(\mbox{\boldmath $m \cdot r$})
              \mbox{\boldmath $r$} - r^2\mbox{\boldmath $m$}]/r^5.  
\label{eq:BDIP}
\end{equation}
The electric
field of the moving neutron is obtained from a Lorentz transformation
of the magnetic field of a motionless neutron[11]
\begin{equation}
\mbox{\boldmath $E$} = -\mbox{\boldmath $v\times B$}.
\label{eq:EN}
\end{equation}

Let us use cylindrical coordinates. The $z$-axis is chosen along
the solenoid's axis. Hence, in this experiment, the neutron's
magnetic dipole is in the $z$-direction, too.
The interesting quantity
is the component of the electric field of the
moving neutron $(\!\!~\ref{eq:EN})$
which is parallel to
the direction of the solenoid's electric current, namely the 
{\boldmath $\phi$}-direction. Since the 
neutron's velocity {\boldmath $v$}
is parallel to the $z$-axis, the relevant magnetic field component
of $(\!\!~\ref{eq:EN})$ is $B_r$. Now, for an infinite
solenoid, one performs an integration on the cylindrical
surface $S$ of the solenoid, uses $(\!\!~\ref{eq:EN})$,
Gauss theorem and Maxwell equation and obtains
\begin{equation}
\int _S E_\phi ds = -v\int_S B_r ds =
-v\int_V \nabla \cdot \mbox{\boldmath $B$}d^3r 
= 0.
\label{eq:B0}
\end{equation}
This result proves that the solenoid's time-depending electric current
is not affected by the electric field of the moving neutron.

   It is proved in [8] that an essential element of the experiment
is the fact that the neutron's internal state remains constant 
throughout the entire experiment. Indeed, it is shown there that if
the neutron is replaced by a ``classical neutron" whose self
energy may change, then the action is independent of the magnetic
field and a null phase shift is obtained. This point emphasizes
the quantum mechanical nature of the experiment.

   The experimental setup of the neutron-electron interaction
has the following properties:
\begin{itemize}
\item[{1.}] The moving particle travels in a force-free region.
\item[{2.}] The overall force exerted by the neutron on the solenoid's
charges vanishes.
\item[{3.}] The change of the action and of the corresponding
phase emerge from the interaction of
the neutron's magnetic dipole {\boldmath $m$} with the solenoid's
magnetic field {\boldmath $B$}, namely, {\boldmath $m \cdot B$}.
\item[{4.}] The neutron remains in its
quantum mechanical ground state throughout the experiment.
\item[{5.}] If the neutron is replaced by a``classical neutron"
whose self energy can be changed
then the action becomes independent of the magnetic field
{\boldmath $B$} and the phase shift disappears.
\end{itemize}
These items are called hereinafter properties 1,...,5, respectively.

   The physics examined here is the phase shift and the
corresponding interference pattern. Examining these phenomena,
one realizes that property 1 above is not essential for the
phase shift effect. It is used only as a convincing proof of the quantum
mechanical nature of the results. Further aspects of the
force-free region are discussed in the last section of this
work.

 In the
rest of this work it is proved that the same kind of
2-body interaction as well as properties 1-5 are found in
an experiment[9] which demonstrates the magnetic AB effect[6,7]. 
In other words, it is proved below that the neutron's interference
experiment[5] confirms the same physical principles as standard
experiments proving the validity of the magnetic AB effect. For
this end, let us start with a brief description of this effect.

\vglue 0.666666in
\noindent
{\bf 3. The Magnetic Aharonov-Bohm Effect}
\vglue 0.33333in

   Consider an infinitely long cylindrically shaped
permanent magnet which is fixed in the
laboratory and its axis coincides with the $z$-axis. Let $R_m$ and
$\Phi $ denote the radius of the magnet's cross section and
its magnetic flux, respectively. An electron moves in the
positive direction of the $y$-axis. At a point $y=-Y_0$,
the electron's wave function is split into two subpackets, 
$\psi _L$ and $\psi _R$, which
continue to move parallel to the $y$-axis along the lines
$x=\pm a,\;z=0$ and pass on the left
and right hand sides of the magnet, respectively. Thus,
the single particle wave function is
\begin{equation}
\psi _e = \psi _L + \psi _R.
\label{eq:PSILR}
\end{equation}
Later, the two subpackets interfere on a screen $S$ (see fig. 1).
In some cases below, this electron is called the travelling electron.

   Let us analyze the influence of the permanent magnet on the
electron's interference pattern. The electron's nonrelativistic
Lagrangian is (see [11], p. 46)
\begin{equation}
L = \frac{1}{2}mv^2 - eV + e \mbox{\boldmath $v\cdot A$},
\label{eq:LAGE}
\end{equation}
where {\boldmath $v$} denotes the electron's velocity and $V$, {\boldmath $A$}
denote the electromagnetic scalar and vector potentials,
respectively.

   In the experiment discussed here, the scalar potential of the
magnet $V$ vanishes. The magnetic field is confined to the inner
part of the magnet, namely to a region where $r < R_m$. Therefore,
the electron travels in a force free region and the first term
of $(\!\!~\ref{eq:LAGE})$ is a constant of the motion. 
Using cylindrical coordinates, one finds that the components
of {\boldmath $A$} at $r>R_m$ are 
\begin{equation}
A_r = A_z = 0,\; A_\phi = \Phi/2\pi r
\label{eq:AOUT}
\end{equation}
where $\Phi $ denotes the solenoid's magnetic flux.
(The validity of this relation is easily verified. 
Using the cylindrical symmetry of the magnet, one takes the
integral along a circle of radius $r$:
$\oint \mbox{\boldmath $A\cdot $}d\mbox{\boldmath $l$}=
\int curl \mbox{\boldmath $A\cdot $}d\mbox{\boldmath $s$}=
\int \mbox{\boldmath $B\cdot $}d\mbox{\boldmath $s$}=\Phi $.)
Hence, the action difference $\Delta I$ between $\psi _L$
and $\psi _R$, associated with the permanent magnet,
is obtained from the substitution of 
$(\!\!~\ref{eq:AOUT})$ into the last term of the Lagrangian
$(\!\!~\ref{eq:LAGE})$. A
straightforward calculation yields (see [6], p. 487)
\begin{eqnarray}
\Delta I      & = & 
          e\int _{-\infty}^\infty \mbox{\boldmath $v\cdot A$}(x=a)dt -
          e\int _{-\infty}^\infty \mbox{\boldmath $v\cdot A$}(x=-a)dt 
\nonumber \\
              & = & e\int _{-\infty}^\infty [A_y(x=a) - A_y(x=-a)]dy 
\nonumber \\
              & = & e\Phi.
\label{eq:DALPHA}
\end{eqnarray}

   This outcome is associated with a phase shift that affects the
interference pattern of the electron. The quantum mechanical
foundations of the results are explained in the following section,
where the role of the quantized state of the magnet is emphasized.

\vglue 0.666666in
\noindent
{\bf 4. The Analogy Between the Polarized Neutron Experiment and 
\newline
$\;\;\;\;\;\;$ the Magnetic Aharonov-Bohm Effect}
\vglue 0.33333in

   Let us analyze the magnetic AB experiment whose principles are 
described above. The travelling electron interacts with the
magnet which is made of neutral atoms, each of which has an
intrinsic magnetic moment. Here, the 2-body interaction takes the
form of $(\!\!~\ref{eq:LLL})$, where the subscript
$n$ is replaced by $A$ which denotes a magnetic atom.
The corresponding 2-body Hamiltonian
and the Schrodinger equation take the form of $(\!\!~\ref{eq:SCH})$.
Thus, one realizes that the underlying 2-body interaction of
the neutron experiment [5] is the same as that of the magnetic
AB effect[6,7,9]. It is proved later that the 2-body interaction
{\boldmath $m\cdot B$}
of $(\!\!~\ref{eq:LAGN})$ equals the corresponding term 
$e${\boldmath $v\cdot A$} of $(\!\!~\ref{eq:LAGE})$.

   On top of that, it is shown here that the experimental
setup designed for measuring the magnetic AB effect
is endowed with the five properties 1...5 of the neutron
experiment. In order to do that, 
one should find the correspondence between
elements of the neutron experiment[5] and those of the magnetic
AB effect. Each experiment consists of a single particle
(a neutron in [5] and an electron in the magnetic 
AB experiment) interacting
with a multitude of other particles (the electrons which make the
solenoid's current and its magnetic field
in [5], and the magnetized atoms which make the
permanent magnet in the magnetic AB case). The linearity of
electrodynamics enables one to write the overall
interaction as a sum of two body
interactions. Thus, the neutron interference experiment[5] is 
based on the interaction of the neutron with an electron moving
along the solenoid's wire. Similarly, 
in the magnetic AB effect, the travelling
electron interacts with a magnetized neutral atom.

   The foregoing discussion shows that the two experiments are
based on a two body interaction of an electron with an 
electrically neutral particle having a nonvanishing magnetic
moment. It follows that the two experiments have an 
intrinsic similarity. In the following analysis, subscripts $e$
and $m$ denote quantities pertaining to the electron and the
magnetic particle, respectively.
Let us prove that the magnetic
AB experiment satisfies properties 1-5. 

   Evidently, the travelling 
electron moves in a field-free region. Hence,
properties 1,2 hold.

   In order to prove property 3, one has to compare the
neutron's Lagrangian $(\!\!~\ref{eq:LAGN})$ with the electron's
Lagrangian $(\!\!~\ref{eq:LAGE})$. 
It turns out that property 3 depends
on the validity of the following relation 
\begin{equation}
\mbox{\boldmath $B$}_e \mbox{\boldmath $\cdot m$} =
e\mbox{\boldmath $v$}_e \mbox{\boldmath $\cdot A$}_m
\label{eq:JABM}
\end{equation}
where $e$ is the electronic charge ($e$ has a negative numerical
value). This relation is proved by means of a direct 
calculation. The origin of coordinates is at the location of the magnet
whose moment is in the $z$-direction
and the electron moves in the $y$-direction.
The left hand side of $(\!\!~\ref{eq:JABM})$,
is calculated first.
Let $R_s$ denote the solenoid's radius and the
electron is at a point $x=R_s,\;y=0,\;z=Z_0$.
At the origin, the $x$-component
of the electric field of the electron is $-e\sin \theta \cos \phi /r^2$.
Using {\boldmath $B$}$=${\boldmath $v\times E$} (which is an
analogue of $(\!\!~\ref{eq:EN})$), and the $y$-direction
of the electron's velocity, one finds that the left hand side
of $(\!\!~\ref{eq:JABM})$ equals $mev\sin \theta \cos \phi/r^2$
(where $m$ denotes the strength of the magnetic dipole and
$\cos \phi = 1$).
Now, let us turn to the right hand side of $(\!\!~\ref{eq:JABM})$.
Using spherical polar coordinates, one finds that
the components of the vector potential of a magnetic dipole whose moment is 
in the $z$-direction are[10]
$A_r=A_\theta =0,\;A_\phi =m\sin \theta /r^2$. Since
the electron's velocity 
$v_y = v\cos \phi$, one finds that the
right hand side of $(\!\!~\ref{eq:JABM})$ equals
$mev\sin \theta \cos \phi/r^2$, too.  
Hence, relation $(\!\!~\ref{eq:JABM})$ holds.

   This calculation proves that the two body interaction of these
experiments are the same. Hence, property 3 holds also for the 
magnetic AB effect. (As a matter of fact, relation $(\!\!~\ref{eq:JABM})$
can be regarded as an extension of a well known 
charge-potential relation of
electrostatics of two charges $q_1V_2=q_2V_1$.)

   The validity of property 4, namely the assumption that the
permanent magnet remains in its ground state throughout the
experiment, is generally taken for granted. This approach
is justified here
by an order of magnitude evaluation
of the interactions involved. Here the interaction of a magnetic
atom with its neighbours has to be compared with its interaction
with the magnetic field {\boldmath $B$}$=${\boldmath $v\times E$}
of the travelling electron.
The inter-atomic distance is
of the order of $10^{-8}$ cm whereas the distance between the
travelling electron and the magnet is about $10^{-4}$ cm[9].
Hence, a comparison with atomic field shows that
the distance-depending factor of the magnetic field
of the travelling electron is weaker by $10^{-8}$. 
Therefore, since the transition probability is proportional
to the square of the ratio of the interactions[12], one
concludes that the transition probability is less than 
$10^{-16}$. In [9], the number of magnetic atoms is about $10^{12}$.
For this reason, the permanent magnet is regarded as 
an inert object whose state is not affected by the fields of
the travelling electron.
It follows that property 4 is confirmed for the magnetic 
AB effect. (It is interesting to note
that in an evaluation of the magnetic
AB effect, the Lagrangian of the magnet's constituents
has to be added to $(\!\!~\ref{eq:LAGE})$. Its omission is justified
only after it is proved that it behaves as an inert object
throughout the experiment.)

The validity of property 5 is examined in a system where the magnet
is replaced by a classical device. This device 
is a cylindrical solenoid which consists of
2 helixes, each of which is a pipe made of an insulating material.
A charged liquid flows frictionlessly along the pipes. 
The electric field of this charge is screened by a static
charge of opposite sign which is spread uniformly on the outer
side of the pipes. The charged fluid
ascends in one pipe and descends in the other. Thus, the mean current
flows in the $(x,y)$ plane. Let {\boldmath $j$} denote the solenoid's 
current ($j=nI$ where $n$ denotes the number of loops per unit
length and $I$ is the ordinary electric current). Hence, the
solenoid's magnetic flux is[13]
\begin{equation}
\Phi = 4\pi ^2R_s^2j,
\label{eq:FLUX}
\end{equation}
where $R_s$ denotes the solenoid's radius.
The Lagrangian of the system is (see [11], p. 46)
\begin{equation}
L = \frac{1}{2}MV^2 + \frac{1}{2}mv^2 + e \mbox{\boldmath $v\cdot A$}
\label{eq:LAGC}
\end{equation}
where $M$ and $V$ denote the mass and velocity of the charged liquid,
respectively and the other terms refer to the travelling electron.
(In this experiment the electric field of the
solenoid vanishes at the outer region of the pipes and the term
depending on the scalar potential is deleted from 
$(\!\!~\ref{eq:LAGC})$. Similarly, motionless parts of the solenoid are
also omitted from this expression.)

   Let us use $(\!\!~\ref{eq:LAGC})$ and calculate the action
for an electron passing on the right hand side of the solenoid
along a line $x=a,\; -\infty <y<\infty,\;z=0$. As mentioned
above, the second term is independent of the magnetic field.
The contribution
of the last term is one half of $(\!\!~\ref{eq:DALPHA})$, namely
\begin{equation}
\Delta I_1 = e\Phi/2.
\label{eq:DELTAI1}
\end{equation}

   The quantity $\frac {1}{2}MV^2$ of $(\!\!~\ref{eq:LAGC})$ changes
during the experiment, due to the force exerted by the 
electric field of the travelling
electron on the charged liquid. Let $\Delta W=M(V^2 - V_0^2)/2$
denote the energy variation. Using cylindrical coordinates and Maxwell
equations, one integrates the power associated with
the electric field of the travelling electron and finds
\begin{eqnarray}
\Delta W & = & \int j\mbox{\boldmath $E \cdot $}
d\mbox{\boldmath $l$}\;dz\;dt
\nonumber \\
         & = & j\int (\;curl \mbox{\boldmath $E$})
 \mbox{\boldmath $\cdot $}
d\mbox{\boldmath $s$}\;dz\;dt
\nonumber \\
         & = & -j\int B_z\;ds\;dz
\label{eq:DELTAW}
\end{eqnarray}
where $d${\boldmath $l$} is a line element in the $\phi$-direction
and $B_z$ is the $z$-component of
the magnetic field of the travelling electron. This
electron moves in the $y$-direction. Hence,
\begin{equation}
B_z = (\mbox{\boldmath $v\times E$})_z = \frac {vae}{(a^2+y^2+z^2)^{3/2}}.
\label{eq:BZ}
\end{equation}

The calculation is carried out first for a very thin solenoid. In
this case, $B_z$ of the travelling electron
is assumed to be uniform at a cross section
of the solenoid
$z=const.$ and the integration on $ds$ boils down
to a multiplication by $\pi R^2_s$. Substituting
$(\!\!~\ref{eq:BZ})$ into $(\!\!~\ref{eq:DELTAW})$ and performing
the integration on $z$[14], one obtains
\begin{equation}
\Delta W = -\frac {2\pi R_s^2avej}{a^2 + y^2}.
\label{eq:D1}
\end{equation}

   Integrating $(\!\!~\ref{eq:D1})$ on the time, one finds the
variation of the action $\Delta I_2$ which emerges from the first term of
$(\!\!~\ref{eq:LAGC})$. Using $v\;dt=dy$
and $(\!\!~\ref{eq:FLUX})$, one obtains
\begin{equation}
\Delta I_2 = -2\pi R_s^2ej\int _{-\infty}^\infty
\frac {a\;dy}{a^2+y^2} = -2\pi^2R_s^2ej =
-e\Phi/2.
\label{eq:DELTAI2}
\end{equation}
 
This result is independent of the impact parameter $a$. Hence, it
also holds for any solenoid, since the latter can be regarded as
an assembly of very thin solenoids.

Adding $(\!\!~\ref{eq:DELTAI1})$ and $(\!\!~\ref{eq:DELTAI2})$, one
obtains a null result. This outcome proves property 5 for the
magnetic AB effect.

\vglue 0.666666in
\noindent
{\bf 5. Topological Features and the Magnetic AB Effect}
\vglue 0.33333in

   The foregoing discussion 
casts a new light on the topological features
of the magnetic AB effect[6,7], where it is required that the
travelling electron should move in a nonsimply connected field-free
region. The proof of item 5 shows that the effect {\em disappears} if
the magnet is replaced by an equivalent 
classical device (which conserves the
nonsimply connectedness of the field-free region). At this point
one may conclude that the nonsimply connectedness is at most
a {\em necessary} (but not {\em sufficient}) 
condition for the magnetic AB effect.

   This matter can be analyzed from another point of view.
A decomposition of the system's interaction into a sum of two
body interactions shows that the nonsimply connectedness is just
a mathematical feature found in one and only one
of two alternative calculations.
Indeed, relation $(\!\!~\ref{eq:JABM})$ shows that the
interaction of the travelling electron with the vector potential
of the magnetic atom 
$e\mbox{\boldmath $v$}_e \mbox{\boldmath $\cdot A$}_m$
can be replaced by 
the interaction of the magnetic atom with the magnetic field of
the travelling electron
$\mbox{\boldmath $B$}_e \mbox{\boldmath $\cdot m$}$. A summation
of the vector potential of
all magnetic atoms at the position {\boldmath $r$}$_{(e)}$
of the travelling electron, 
yields the electron's interaction with the 
vector potential of the entire magnet $(\!\!~\ref{eq:AOUT})$.
\begin{equation}
L_{int} = e\mbox{\boldmath $v$}_e \mbox{\boldmath $\cdot$}
[(\Sigma \mbox{\boldmath $A$}_{(m)i}
(\mbox{\boldmath $r$}_{(e)})] =
e\mbox{\boldmath $v$}_e \mbox{\boldmath $\cdot A$}.
\label{eq:ABP}
\end{equation}
In this picture, one obtains
an electron moving in a field-free region which is multiply
connected. (Here and below, the subscripts $m,e$ are put
in brackets, in order to be distinguished from the summation
index $i$.)

   On the other hand, one may perform the summation on
the magnetic atoms {\boldmath $m$}$_i$
interacting with the magnetic field of the
travelling electron and find for the same physical experiment
\begin{equation}
L_{int} = \mbox{$\Sigma \mbox{\boldmath $B$}_{(e)}
(\mbox{\boldmath $r$}_{(m)i}$) \mbox{\boldmath $\cdot m$}$_i$}.
\label{eq:MYP}
\end{equation}
Here the magnetic moment of each atom interacts with the local
magnetic field of the travelling electron 
{\boldmath $B$}$_e(\mbox{\boldmath $r$}_{(m)i})$
and no multiply connected field-free region exists because the
magnetic field of the travelling electron is nonzero at the
location of the relevant magnetic atoms.

   Evidently, the 2 pictures describing the interaction of
the travelling electron with the source of the magnetic field
are equivalent and one may use either of them.
Since the topological field-free
property does not exist in the second picture, one concludes
that in the magnetic AB effect, topology
has no profound physical meaning.

\vglue 0.666666in
\noindent
{\bf 6. Conclusions}
\vglue 0.33333in

   It is proved in this work that the polarized neutron experiment[5]
and the one demonstrating the magnetic AB effect[9] are based on the
same physical principles. Indeed, the two body interaction of each
of them is an interaction of a charge (an electron) with a neutral
particle having an intrinsic magnetic moment (a neutron in [5] and
a magnetic atom in [9]). Furthermore, the main features of the
experimental setup of the two experiments are the same.

   It is further proved here that the inert nature of the magnetic
neutral component is vital for the nonvanishing phase shift
obtained. This point is a manifestation of
the quantum mechanical nature of
the effect demonstrated 
by the two experiments. Indeed, it is shown in [8]
that if the neutron is replaced by a classical device then the
effect disappears. An analogous property holds for the magnetic AB
effect, as shown in the proof of item 5 in 
the last part of section 4. In section
5 it is proved that topological features of the field-free
region do not have a fundamental physical significance.

   It can be concluded that the quantum mechanical foundations of
the polarized neutron experiment 
and of the magnetic AB effect are based on
the structure of their magnetic dipole constituents that behave
as inert objects. These objects stay in their 
quantum mechanical state throughout the experiment and yield
a nonzero phase shift that affects the interference pattern.

\newpage
References:
\begin{itemize}
\item[{[1]}] C. G. Shull, E. O. Wollan and W. A. Strauser,
Phys. Rev. {\bf 81} (1951) 483.
\item[{[2]}] F. Mezei, Physica {\bf 137B-C} (1986) 295.
\item[{[2]}] F. Mezei, Physica {\bf 151B-C} (1988) 74.
\item[{[4]}] J. Summhammer et al.,
Phys. Rev. Lett. {\bf 75} (1995) 3206.
\item[{[5]}] W. -T Lee, O. Motrunich, B. E. Allman and S. A. Werner,
Phys. Rev. Lett. {\bf 80} (1998) 3165.
\item[{[6]}] Y. Aharonov and D. Bohm, Phys. Rev. {\bf 115} (1959)
485.
\item[{[7]}] Y. Aharonov and D. Bohm, Phys. Rev. {\bf 123} (1961)
1511.
\item[{[8]}] E. Comay, Phys. Lett. {\bf A250} (1998) 12
(stored as quant-ph/9906058 at quant-ph@xxx.lanl.gov).
\item[{[9]}] A. Tonomura et al., Phys. Rev. Lett. {\bf 56},
(1986) 792.
\item[{[10]}]  J. D. Jackson, {\em Classical Electrodynamics,}
second edition (John Wiley, New York, 1975). p. 182.
\item[{[11]}] L. D. Landau and E. M. Lifshitz, {\em The Classical
Theory of Fields} (Pergamon, Oxford, 1975). p. 62.
\item[{[12]}] L. I. Schiff, Quantum Mechanics, (Mc Graw-Hill,
New York, 1955). p. 197.
\item[{[13]}] E. M. Purcell, Electricity and Magnetism (McGraw-Hill,
New York, 1985). p. 228.
\item[{[14]}] I. S. Gradshteyn and I. M. Ryzhik, Tables of
Integrals, Series and Products (Academic Press, Boston, 1994).
p. 101 (item 5).

\end{itemize}

\newpage
Figure captions

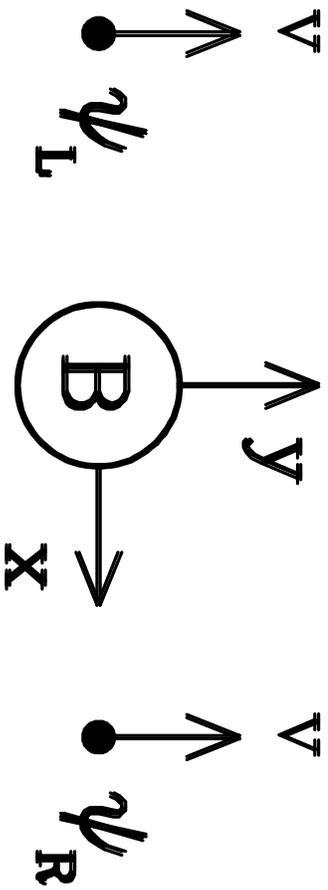
Fig. 1:

The magnetic AB effect. Two subpackets of an electron,
$\psi _L$ and $\psi _R$, move in the $(x,y)$ plane.
The subpackets move parallel to the $y$-axis and pass on
either side of a magnet. The axis of the infinitely long magnet
coincides with the $z$-axis and its magnetig field
{\boldmath $B$} is confined to the magnet's inner part.
Later, the subpackets interfere on the screen $S$.

\end{document}